\documentclass[12pt,a4paper]{article}

\usepackage[truedimen,margin=30mm]{geometry} 

\usepackage{amssymb}
\usepackage{amsmath}
\usepackage[hiresbb]{graphicx}
\usepackage{amsthm}   %This is necessary for "proof"
\usepackage{authblk}   %This is necessary for author descriptions
\usepackage{lscape}   %This is necessary for "landscape"
\usepackage{url}
\usepackage{multirow}
\usepackage{natbib}

\usepackage{setspace}

\usepackage{times}
\usepackage{dsfont}
\usepackage{subcaption}
\usepackage{here}
\usepackage{booktabs, threeparttable}

\makeatletter
\def\mojiparline#1{
    \newcounter{mpl}
    \setcounter{mpl}{#1}
    \@tempdima=\linewidth
    \advance\@tempdima by-\value{mpl}zw
    \addtocounter{mpl}{-1}
    \divide\@tempdima by \value{mpl}
    \advance\kanjiskip by\@tempdima
    \advance\parindent by\@tempdima
}
\makeatother
\def\linesparpage#1{
    \baselineskip=\textheight
    \divide\baselineskip by #1
}

\allowdisplaybreaks[4]

\title{ {\bf Posterior Quantification of Borrowing from Multiple Historical Control Data in Bayesian Dynamic Borrowing Methods: A Scoping Review } }

\author[1]{Tomohiro Ohigashi}
\author[2,3]{Wataru Murasaki}
\author[4]{Masahiko Gosho}

\affil[1]{Department of Information and Computer Technology, Faculty of Engineering, Tokyo University of Science, Tokyo, Japan}
\affil[2]{Graduate School of Comprehensive Human Sciences, University of Tsukuba, Ibaraki, Japan}
\affil[3]{Department of Biostatistics, Tsukuba Clinical Research \& Development\\ Organization, University of Tsukuba, Ibaraki, Japan}
\affil[4]{Department of Biostatistics, Institute of Medicine, University of Tsukuba, Ibaraki, Japan}

\date{}

\begin{document}
\linesparpage{25}
\allowdisplaybreaks[4]
\begin{singlespace}
\maketitle
\end{singlespace}

\vspace{-1cm}
\begin{center}
{\large\bf Abstract}
\end{center}
Bayesian dynamic borrowing methods incorporate historical control data into current clinical trial analyses while allowing the degree of borrowing to depend on the compatibility between historical and current data. Although many methods have been proposed, the degree of borrowing is often difficult to interpret, especially when multiple historical control sources are available. This scoping review focuses on posterior quantification of borrowing from multiple historical controls. We discuss overall borrowing summaries based on effective historical sample size, together with method-specific source-level summaries of borrowing, information contribution, or compatibility arising from power priors, unit information priors, multisource exchangeability models, Dirichlet process mixture models, and potential bias models. We distinguish posterior borrowing measures from quantities describing prior information allocation or source-specific conflict. Two case studies, one with a binary endpoint and one with a continuous endpoint, illustrate that methods with broadly similar posterior treatment effect estimates may differ in both the overall amount and source-specific pattern of borrowing. These examples show that large overall borrowing may reflect selective borrowing from compatible historical sources rather than uniform borrowing from all sources. We recommend reporting treatment effect estimates together with overall and source-specific borrowing summaries, when available, to improve transparency in posterior inference.


\section{Introduction\label{sec:intro}}
The 21st Century Cures Act, passed in the U.S. in 2016, led the Food and Drug Administration (FDA) to provide guidelines for using existing data, with relevant infrastructure being developed to support this initiative.
In clinical trials for rare diseases and those involving children, it is often difficult to enroll a sufficient number of participants, making the effective use of historical or external data desirable \citep{limMinimizingPatientBurden2018}.
Regulatory interest in the formal incorporation of existing information has also been reflected in recent guidance on Bayesian clinical trials. 
The FDA's draft guidance on the use of Bayesian methodology in clinical trials of drug and biological products discusses the appropriate use of Bayesian methods in submissions for drugs and biologics, including their use to support primary inference in trials intended to provide evidence of effectiveness and safety \citep{fdaBayesianMethodologyClinical2026}. 
These regulatory developments highlight the importance of transparent approaches to specifying, quantifying, and interpreting the contribution of historical or external data in clinical trial analyses.

\citet{pocockCombinationRandomizedHistorical1976} focused on approaches to incorporate historical/external data, referred to hereafter as historical control data, into the control arms of current randomized controlled trials (RCTs).
This approach is known as a hybrid controlled trial design and has been adopted in real-world clinical studies \citep{baetenAntiinterleukin17AMonoclonalAntibody2013}.
Among the analytical methods for such trials, Bayesian dynamic borrowing methods have received considerable attention \citep{vieleUseHistoricalControl2014}.
In the context of Bayesian dynamic borrowing, the discrepancy between the information provided by historical control data and that provided by the control group data from the target RCT is referred to as prior--data conflict.
The degree of borrowing from the historical control data is then adjusted according to the extent of this conflict.
These methods address prior--data conflict by imposing various assumptions on the relationship between the parameters of the historical control data and those of the current control data, and by incorporating these assumptions into the prior and likelihood.
Although several Bayesian dynamic borrowing methods have been proposed, their sensitivity to prior--data conflict and the resulting degree of information borrowing can differ substantially across methods.
Moreover, when multiple historical control sources are available, prior--data conflict may arise at different levels.
We refer to global conflict between the current control and the collection of historical controls as between-conflict, and local conflict attributable to particular historical control sources as within-conflict.
This distinction can help clarify whether borrowing is reduced overall or selectively from specific incompatible sources.

Consequently, examining the posterior treatment effect estimate and its credible interval alone may provide only a partial view of a Bayesian dynamic borrowing analysis.
Global summaries such as the effective historical sample size (EHSS) \citep{hobbsAdaptiveAdjustmentRandomization2013, wiesenfarthQuantificationPriorImpact2020, bennettNovelEquivalenceProbability2021} are useful for describing the overall contribution of historical data. 
Because EHSS summarizes overall borrowing, it can help characterize reductions in overall borrowing that may occur in the presence of between-conflict.
When multiple historical control sources are incorporated, it can also be informative to examine how borrowing is distributed across individual sources and how this pattern relates to prior--data conflict.
These considerations motivate the review of approaches to quantify both overall and source-specific borrowing in Bayesian dynamic borrowing analyses.

This article reviews and illustrates approaches for quantifying information borrowing in Bayesian dynamic borrowing analyses, with particular emphasis on source-specific borrowing when multiple historical control sources are available.
Rather than focusing solely on the construction of individual borrowing models or their operating characteristics, we highlight how the resulting degree of borrowing can be summarized, visualized, and interpreted.
We consider both binary and continuous outcomes and use two case studies to illustrate how overall and source-specific borrowing measures can help characterize the behavior of different dynamic borrowing methods in the presence of prior--data conflict.

The remainder of this paper is organized as follows.
In Section \ref{sec:method}, we review Bayesian dynamic borrowing methods.
In Section~\ref{sec:quantify}, we review approaches for quantifying information borrowing and prior--data conflict, including overall and source-specific borrowing measures.
In Sections \ref{sec:binary} and \ref{sec:cont}, we illustrate the methods through two applications to real datasets.
In Section \ref{sec:recomm}, we discuss practical recommendations for reporting and interpreting Bayesian dynamic borrowing analyses.
We conclude in Section \ref{sec:dis} with a discussion of implications and remaining challenges.

\section{Dynamic borrowing methods\label{sec:method}}

\subsection{Meta-analytic approach\label{sec:MA}}

The meta-analytic predictive (MAP) approach is a widely used Bayesian framework for incorporating historical control data into the analysis of a current trial \citep{neuenschwanderSummarizingHistoricalInformation2010, schmidliRobustMetaanalyticpredictivePriors2014}.
Let $\theta_{\rm CC}$ denote the parameter of interest for the current control, and let $\theta_{{\rm H}_k}$ denote the corresponding parameter for the $k$-th historical control, $k=1,\ldots,K$.
The MAP approach models the historical and current control parameters as arising from a common population distribution.
In contrast to the meta-analytic combined approach, which jointly analyzes historical and current control data, the MAP approach is conducted in two stages.
First, a meta-analytic model is fitted to the historical control data to derive the predictive distribution of the parameter in a current trial.
This predictive distribution is used as the MAP prior for $\theta_{\rm CC}$.
Second, the MAP prior is updated using the current control data.
A common exchangeability model assumes
\[
  \theta_j \mid \mu_{\rm MA}, \tau_{\rm MA}
  \sim {\rm N}\left(\mu_{\rm MA}, \tau_{\rm MA}^2\right),
  \quad j = {\rm CC}, {\rm H}_1,\ldots,{\rm H}_K,
\]
where $\mu_{\rm MA}$ is the overall mean and $\tau_{\rm MA}^2$ is the between-trial variance.
The degree of borrowing depends on the heterogeneity parameter $\tau_{\rm MA}$: smaller values imply stronger borrowing, whereas larger values yield a more diffuse MAP prior and reduce the influence of historical information.

Robust MAP priors have been proposed to reduce the influence of historical information when prior--data conflict is present \citep{schmidliRobustMetaanalyticpredictivePriors2014}.
A common robustification strategy combines the MAP prior with a weakly informative robust component,
\[
  \pi(\theta_{\rm CC}) = (1-w_{\rm R}) \pi_{\rm MAP}(\theta_{\rm CC}) + w_{\rm R} \pi_{\rm R}(\theta_{\rm CC}),
\]
where $\pi_{\rm R}$ is the robust component and $w_{\rm R}$ is its mixture weight.
This construction reduces reliance on the MAP prior when the current data are inconsistent with the historical information.

A more flexible extension is the Dirichlet process mixture MAP (DPM-MAP) approach \citep{hupfBayesianSemiparametricMetaanalyticpredictive2021}.
Instead of using a single parametric distribution for between-trial heterogeneity, DPM-MAP models latent heterogeneity among historical and current control parameters through a Dirichlet process mixture:
\[
\begin{aligned}
\theta_j &= \mu + \delta_j, \\
\delta_j &\sim {\rm N}\left(0, \tau_j^2\right), \\
\tau_j \mid G &\sim G, \\
G &\sim {\rm DP}\left(\alpha_{\rm DP}, G_0\right),
\end{aligned}
\]
for $j = {\rm CC}, {\rm H}_1,\ldots,{\rm H}_K$.
Here, $\delta_j$ is a trial-specific deviation, $\tau_j$ is a trial-specific standard deviation parameter, and $G$ is a nonparametric mixing distribution for $\tau_j$.
The distribution $G$ is assigned a Dirichlet process prior, denoted by ${\rm DP}(\alpha_{\rm DP},G_0)$, with concentration parameter $\alpha_{\rm DP}$ and base distribution $G_0$.
By allowing flexible modeling of trial-specific variance components, DPM-MAP can represent more flexible heterogeneity structures than the conventional MAP model with a single between-trial variance.
In practical computations, the infinite mixture representation is often approximated by a finite truncation, such as a truncated stick-breaking process \citep{ishwaranApproximateDirichletProcess2002, ishwaranExactApproximateSum2002}.
Although DPM-MAP uses a Dirichlet process mixture to model between-trial heterogeneity, it does not define source-specific borrowing based on shared clustering of the historical and current control parameters.

\subsection{Power prior\label{sec:PP}}

The power prior incorporates historical control data by discounting the historical likelihood \citep{chenPowerPriorDistributions2000}.
Let $D_{{\rm H}_k}$ and $D_{\rm CC}$ denote the $k$-th historical control data and the current control data, respectively.
The historical and current control data are modeled through a common parameter, $\theta$, while potential discrepancies are handled by discounting the historical likelihoods.
For multiple historical control sources, the power prior can be written as:
\[
  \pi(\theta \mid D_{{\rm H}_1},\ldots,D_{{\rm H}_K},\boldsymbol{\gamma})
  \propto
  \pi_0(\theta)
  \prod_{k=1}^K
  L(\theta \mid D_{{\rm H}_k})^{\gamma_{{\rm H}_k}},
\]
where $\pi_0(\theta)$ is the initial prior, $L(\theta \mid D_{{\rm H}_k})$ is the likelihood contribution from the $k$-th historical control, and $\gamma_{{\rm H}_k} \in [0,1]$ is the corresponding power parameter.
When $\gamma_{{\rm H}_k}=1$, the $k$-th historical source is fully borrowed, whereas when $\gamma_{{\rm H}_k}=0$, it is completely discounted.

In the fully Bayesian formulation, the power parameters are assigned prior distributions and estimated jointly with the model parameters, allowing the amount of borrowing to adapt to the compatibility between historical and current data.
For multiple historical controls, \citet{banbetaModifiedPowerPrior2019} proposed modified power prior methods with source-specific power parameters.
Among these, the dependent modified power prior (DMPP) assumes a hierarchical structure,
\[
  \gamma_{{\rm H}_1},\ldots,\gamma_{{\rm H}_K}
  \mid a_{\rm PP}, b_{\rm PP}
  \sim
  {\rm Beta}\left(a_{\rm PP}, b_{\rm PP}\right),
\]
where $a_{\rm PP}$ and $b_{\rm PP}$ control the distribution of borrowing across historical sources.
Equivalently, the beta distribution can be parametrized by its mean $\mu_{\rm PP}=a_{\rm PP}/(a_{\rm PP}+b_{\rm PP})$ and variance $\tau_{\rm PP}^2 = \frac{\mu_{\rm PP}(1-\mu_{\rm PP})} {a_{\rm PP}+b_{\rm PP}+1}$.
This hierarchical formulation allows the borrowing weights to vary by source while sharing information across historical controls.
The posterior distribution of $\gamma_{{\rm H}_k}$ provides a natural summary of source-specific borrowing, because each power parameter directly controls the contribution of the corresponding historical likelihood.

\subsection{Potential bias models\label{sec:PBM}}

The potential bias model (PBM) represents the discrepancy between historical and current control parameters through a source-specific bias term:
\[
  \theta_{{\rm H}_k} = \theta_{\rm CC} + \beta_{{\rm H}_k},
  \quad k=1,\ldots,K,
\]
where $\beta_{{\rm H}_k}$ denotes the potential bias between the $k$-th historical control and current control.
When $\beta_{{\rm H}_k}$ is close to zero, the corresponding historical source is regarded as compatible with the current control; when it deviates from zero, the historical source differs from the current control.
For a single historical control source, related formulations are often discussed as commensurate priors, which introduce a commensurability structure between the historical and current parameters \citep{hobbsHierarchicalCommensuratePower2011}.

\citet{ohigashiUsingHorseshoePrior2022} proposed assigning a horseshoe prior to the potential bias parameters.
The horseshoe prior is a global--local shrinkage prior that strongly shrinks small signals toward zero while allowing large signals to escape shrinkage \citep{carvalhoHorseshoeEstimatorSparse2010}.
This property is useful in the PBM because most historical sources may be compatible with the current control, while a small number of sources may exhibit prior--data conflict.
The PBM with the horseshoe prior assumes
\[
\begin{aligned}
\beta_{{\rm H}_k} 
  &\sim {\rm N}\left(0, \lambda^2_{{\rm H}_k} \tau^2_{\rm HS}\right), \\
\lambda_{{\rm H}_k} 
  &\sim {\rm C^+}(0,1), \\
\tau_{\rm HS} 
  &\sim {\rm C^+}(0,1),
\end{aligned}
\]
where ${\rm C^+}(0,1)$ denotes the standard half-Cauchy distribution.
The global shrinkage parameter $\tau_{\rm HS}$ controls the overall degree of shrinkage of the potential biases toward zero, whereas the local shrinkage parameter $\lambda_{{\rm H}_k}$ allows the $k$-th historical source to deviate from the current control when needed.
Other shrinkage priors, such as spike-and-slab priors, can also be assigned directly to the potential bias parameters depending on how compatibility and conflict are modeled \citep{ohigashiPotentialBiasModels2025}.

From the perspective of dynamic borrowing, the PBM induces borrowing by shrinking source-specific potential biases toward zero.
When current and $k$-th historical controls are compatible, the posterior distribution of $\beta_{{\rm H}_k}$ tends to concentrate around zero, leading to greater borrowing from that source.
In contrast, when the $k$-th historical control is in conflict with the current control, the local shrinkage parameter can reduce shrinkage for $\beta_{{\rm H}_k}$, limiting the influence of that historical source on inference for the current control.
Thus, the posterior behavior of $\beta_{{\rm H}_k}$ and $\lambda_{{\rm H}_k}$ is directly related to source-specific conflict and borrowing.

\subsection{Unit information prior\label{sec:UIP}}

\citet{jinUnitInformationPrior2021} proposed the unit information prior (UIP), which constructs a prior distribution for the current control parameter using summary statistics from multiple historical controls.
In the UIP, information from multiple historical controls is incorporated through a weighted average of historical estimates and the amount of borrowed information.

For a binary response, let $n_{{\rm H}_k}$ and $y_{{\rm H}_k}$ denote the observed sample size and number of responses in the $k$-th historical control, respectively.
The response rate in the $k$-th historical control is given by
$\hat{p}_{{\rm H}_k} = y_{{\rm H}_k}/n_{{\rm H}_k}$.
The mean of the UIP is defined as a weighted average of the historical response rates $\mu_{\rm UIP} = \sum^K_{k=1} w_{\text{UIP},k} \hat{p}_{{\rm H}_k}$, where $w_{\text{UIP},k}$ denotes the weight assigned to the $k$-th historical control, with
$0 \leq w_{\text{UIP},k} \leq 1$ and $\sum^K_{k=1} w_{\text{UIP},k}=1$.
A larger value of $w_{\text{UIP},k}$ indicates a larger relative contribution of the $k$-th historical control to the prior mean \citep{jinUnitInformationPrior2021}.
The variance of the UIP is defined as $\tau^2_{\rm UIP} = \left\{M \sum^K_{k=1} w_{\text{UIP},k} I_U\left(\hat{p}_{{\rm H}_k}\right) \right\}^{-1}$, where $I_U\left(\hat{p}_{{\rm H}_k}\right)$ denotes the unit information evaluated at the observed response rate of the $k$-th historical control.
The unit information is the observed Fisher information averaged at the unit-observation level.
The parameter $M$ represents the amount of information borrowed from historical controls in units of sample size, and can be assigned a non-informative uniform prior,
${\rm Unif}\left(0, \sum^K_{k=1} n_{{\rm H}_k}\right)$.
Therefore, $M$ can be interpreted as a global borrowing amount from the set of historical controls, whereas $w_{\text{UIP},k}$ represents the relative contribution of each historical control.
\citet{jinUnitInformationPrior2021} assigned a Dirichlet prior,
${\rm Dir}\left(\nu_{{\rm H}_1}, \ldots, \nu_{{\rm H}_K}\right)$,
to the weight vector
$\left(w_{\text{UIP},1}, \ldots, w_{\text{UIP},K}\right)$.
They proposed setting
$\nu_{{\rm H}_k} = {\rm min}\left(1, n_{{\rm H}_k}/n_{\rm CC}\right)$
to reduce the influence of a historical control with a large sample size.
For a binary response, $\mu_{\rm UIP}$ and $\tau^2_{\rm UIP}$ are used to determine the corresponding beta prior for the current control response probability.

\subsection{Multisource-exchangeability model\label{sec:MEM}}

\citet{kaizerBayesianHierarchicalModeling2018} proposed the multisource exchangeability model (MEM), which incorporates multiple supplemental data sources into a primary analysis through Bayesian hierarchical modeling.
The MEM considers possible exchangeability patterns, each specifying which historical controls are exchangeable with the current control.
For each pattern, historical controls regarded as exchangeable with the current control contribute to posterior inference for the current control parameter, whereas non-exchangeable historical controls do not contribute directly to inference for the current control parameter under that pattern.
Let $\Omega$ denote one such exchangeability pattern, and let $\mathcal{O}$ be the collection of patterns considered in the MEM.
The posterior distribution of the current control parameter $\theta_{\rm CC}$ is obtained using Bayesian model averaging as follows:
\[
\pi\left(\theta_{\rm CC} \mid D_{\rm CC}, D_{\rm H}\right)
=
\sum_{\Omega \in \mathcal{O}}
\pi\left(\theta_{\rm CC} \mid D_{\rm CC}, D_{\rm H}, \Omega\right)
\pi\left(\Omega \mid D_{\rm CC}, D_{\rm H}\right),
\]
where $D_{\rm H}=\{D_{{\rm H}_1},\ldots,D_{{\rm H}_K}\}$.

For the $k$-th historical control, the degree of borrowing can be summarized by the posterior probability that it is exchangeable with the current control.
Let $p_{{\rm EX},k}$ denote the posterior probability that the $k$-th historical control is exchangeable with the current control.
A larger value of $p_{{\rm EX},k}$ indicates stronger posterior support for exchangeability between the $k$-th historical control and the current control, and thus greater borrowing from that historical control.
Conversely, a smaller value of $p_{{\rm EX},k}$ indicates weaker evidence of exchangeability and reduced borrowing from that historical control.
In this sense, the MEM provides source-specific posterior summaries of borrowing through posterior exchangeability probabilities.

\subsection{Dirichlet process mixture models\label{sec:DPM}}

\citet{ohigashiNonparametricBayesianApproach2025} proposed Dirichlet process mixture (DPM) and dependent DPM (DDPM) models for dynamic borrowing of multiple historical control data.
These methods use latent clustering to determine whether each historical control is compatible with the current control.
Borrowing occurs when a historical control and the current control are assigned to the same mixture component.

Let $\theta_{{\rm H}_k}$ and $\theta_{\rm CC}$ denote the parameters of interest for the $k$-th historical control and the current control, respectively.
In the DPM model, these parameters are assumed to arise from a common Dirichlet process:
\begin{eqnarray*}
\theta_j \mid G &\sim& G, \quad
j \in \{{\rm H}_1,\ldots,{\rm H}_K, {\rm CC}\}, \nonumber\\
G &\sim& {\rm DP}(M,G_0),
\end{eqnarray*}
where $M$ is the concentration parameter and $G_0$ is the base measure.
Because a realization from a Dirichlet process is discrete, a historical control and the current control may share the same mixture component.
This shared clustering provides a mechanism for borrowing information from compatible historical controls.
Let $c_{{\rm H}_k}$ and $c_{\rm CC}$ denote the latent cluster indicators for the $k$-th historical control and the current control, respectively.
\citet{ohigashiNonparametricBayesianApproach2025} defined the similarity and borrowing index (SBI) for the $k$-th historical control as the posterior probability of shared clustering ${\rm SBI}_k
=
{\rm Pr}\left(c_{\rm CC}=c_{{\rm H}_k} \mid D_{\rm CC},D_{\rm H}\right)$.
The SBI can be interpreted as a source-specific posterior summary of both similarity and borrowing.
A larger value of ${\rm SBI}_k$ indicates higher posterior probability that the $k$-th historical control and the current control are compatible, supporting greater borrowing from that historical control.

The DDPM \citep{quintanaDependentDirichletProcess2022} extends the DPM model by introducing dependent random probability measures for the historical and current controls.
In the DDPM, historical controls are modeled using $G_{\rm H}$, whereas the current control is modeled using $G_{\rm CC}$.
These random measures are dependent through common atoms and dependent mixture weights.
This construction allows DDPM to reduce borrowing when the current control is likely to belong to a different mixture component from the historical controls.
Thus, in both DPM and DDPM, the SBI provides a source-specific posterior index of similarity and borrowing through shared clustering, while DDPM allows more flexible reduction of borrowing from historical controls that conflict with the current control.

\section{Quantifying borrowing and conflict\label{sec:quantify}}

In Bayesian dynamic borrowing analyses, the amount of information borrowed from historical controls is not directly observed.
Instead, it is characterized through posterior quantities that summarize the contribution of historical information to inference for the current control parameter.
When multiple historical controls are incorporated, this contribution can be summarized at two levels: overall borrowing from the collection of historical controls and source-specific summaries for individual historical controls.

Borrowing and prior--data conflict are closely related but conceptually distinct. 
The distinction between between-conflict and within-conflict introduced above motivates summarizing borrowing at both the overall and source-specific levels.
A large degree of borrowing generally indicates compatibility between historical and current controls, whereas reduced borrowing may indicate prior--data conflict or a model-based reduction in the influence of historical information.
However, the quantities available for assessing borrowing differ across methods.
Several methods provide direct borrowing parameters, whereas others provide posterior measures of compatibility, exchangeability, clustering, or bias that indirectly characterize borrowing.
Therefore, these quantities should be interpreted according to the borrowing mechanism of each method.
The model components that can be used to summarize source-specific borrowing or compatibility differ across methods.
In contrast, overall borrowing can be summarized across methods using posterior effective sample size (ESS)-based quantities such as EHSS, provided that an appropriate posterior ESS definition is specified.
Table~\ref{tab:borrowing_summaries} summarizes the main borrowing mechanisms, conflict-related model components, and the corresponding overall and source-specific summaries.
The conflict-related components listed in the table should not necessarily be interpreted as direct conflict parameters.
They are model components or derived posterior summaries through which reduced compatibility or prior--data conflict may be reflected.

\begin{table}[htbp]
\centering
\caption{Method-specific borrowing mechanisms, conflict-related model components, and posterior summaries for overall and source-specific borrowing or compatibility in Bayesian dynamic borrowing with multiple historical controls.}
\label{tab:borrowing_summaries}
\footnotesize
\begin{tabular}{p{0.1\textwidth}p{0.2\textwidth}p{0.22\textwidth}p{0.20\textwidth}p{0.22\textwidth}}
\toprule
Method & Borrowing mechanism & Conflict-related component & Overall  & Source-specific \\
\midrule
MAP / robust MAP / DPM-MAP
& Exchangeability and robustification
& Heterogeneity parameters; robust component
& EHSS
& Not directly defined \\

DMPP
& Discounting of historical likelihoods
& Source-specific power parameter $\gamma_{{\rm H}_k}$
& EHSS
& Posterior $\gamma_{{\rm H}_k}$ \\

PBM
& Shrinkage of source-specific potential bias
& Potential bias $\beta_{{\rm H}_k}$ and shrinkage parameters
& EHSS
& Posterior $\beta_{{\rm H}_k}$ and shrinkage parameters as conflict/compatibility summaries \\

UIP
& Construction of an informative prior from unit information
& Not directly defined
& EHSS; posterior $M$ as information incorporated through the UIP
& Posterior \(M w_{{\rm UIP},k}\) as source-specific contribution to the constructed UIP \\

MEM
& Bayesian model averaging over exchangeability patterns
& Inclusion of each historical source in exchangeability patterns
& EHSS
& Posterior exchangeability probability $p_{{\rm EX},k}$ \\

DPM / DDPM
& Borrowing through clustering of historical and current control parameters
& Latent cluster allocation indicators $c_{\rm CC}$ and $c_{{\rm H}_k}$
& EHSS
& Similarity and borrowing index ${\rm SBI}_k$ \\
\bottomrule
\end{tabular}
\end{table}

\subsection{Overall borrowing measures}\label{sec:overall_borrowing}

ESS expresses the amount of statistical information contained in a prior or posterior distribution on the scale of sample size \citep{moritaDeterminingEffectiveSample2008,moritaPriorEffectiveSample2012, neuenschwanderPredictivelyConsistentPrior2020,tamanoiPriorEffectiveSample2024}.
Although ESS is straightforward in certain conjugate settings, its calculation for general prior or posterior distributions depends on the target quantity and the definition of information used. 
Several approaches have been proposed, including moment-matching, curvature-based, and information-ratio-based methods. 
In this review, we calculate EHSS from posterior ESS values obtained using the expected local-information-ratio (ELIR) method \citep{neuenschwanderPredictivelyConsistentPrior2020}.
In Bayesian dynamic borrowing, prior ESS quantifies the information contained in the borrowing prior before the current control data are observed.
Although prior ESS is useful for assessing the informativeness of the prior induced by historical controls, it does not necessarily represent the actual amount of information borrowed after observing the current data.

Because dynamic borrowing adjusts the contribution of historical controls according to their compatibility with the current control, the borrowed information is naturally evaluated using posterior quantities.
Let ${\rm ESS}_{\rm post}$ denote the posterior ESS for the current control parameter after incorporating both historical and current control data, and let $n_{\rm CC}$ denote the current control sample size.
EHSS \citep{hobbsAdaptiveAdjustmentRandomization2013, wiesenfarthQuantificationPriorImpact2020, bennettNovelEquivalenceProbability2021} can be defined as ${\rm EHSS} = {\rm ESS}_{\rm post} - n_{\rm CC}$.
EHSS represents the additional information that cannot be obtained from the current control data alone and can be interpreted as the overall amount of information borrowed from historical controls.

When historical controls are compatible with the current control as a collection, ${\rm ESS}_{\rm post}$ may substantially exceed $n_{\rm CC}$, resulting in a large EHSS. 
Conversely, when between-conflict is present, dynamic borrowing methods may reduce the overall influence of historical information, and ${\rm ESS}_{\rm post}$ approaches $n_{\rm CC}$, resulting in a small EHSS.
Therefore, EHSS provides an interpretable posterior summary of overall borrowing on the sample-size scale.

EHSS should be distinguished from the effective current sample size (ECSS) of
\citet{wiesenfarthQuantificationPriorImpact2020}.
ECSS quantifies prior impact in units of samples from the current data model and is useful for assessing prior--data conflict or for adaptive sample-size decisions.
In this review, we use EHSS as a descriptive posterior borrowing summary, representing the increase in posterior information for the current control parameter after fitting a borrowing model.

\subsection{Source-specific borrowing and compatibility measures}\label{sec:source_specific_borrowing}

When multiple historical controls are available, source-specific summaries are useful for describing how each historical source contributes to inference for the current control and for identifying within-conflict.
The interpretation of these summaries depends on the borrowing mechanism of each method.
Several quantities directly control the contribution of a historical likelihood, whereas others summarize posterior support for compatibility between a historical and the current control.

\noindent\textit{MAP-type approaches.}
For MAP-type approaches, including robust MAP and DPM-MAP, source-specific posterior borrowing summaries are not directly defined in the posterior analysis.
Because the MAP prior is constructed from the collection of historical controls, the posterior contribution of each historical source is not naturally separated after updating with the current control data.

\noindent\textit{Dependent modified power prior.}
For the DMPP, the posterior distribution of the source-specific power parameter $\gamma_{{\rm H}_k}$ directly summarizes the contribution of the $k$-th historical likelihood.
Thus, posterior summaries such as the mean, median, and credible interval of $\gamma_{{\rm H}_k}$ can be used as source-specific borrowing measures.

\noindent\textit{Potential bias models.}
For PBMs, within-conflict is directly represented by the source-specific potential bias parameter $\beta_{{\rm H}_k}$.
Posterior concentration of $\beta_{{\rm H}_k}$ around zero indicates compatibility between the $k$-th historical control and the current control, whereas posterior mass away from zero indicates source-specific conflict.
In PBMs with shrinkage priors, borrowing is induced indirectly by shrinking $\beta_{{\rm H}_k}$ toward zero.
Thus, posterior summaries of $\beta_{{\rm H}_k}$ and its associated shrinkage parameters can be used as source-specific compatibility or conflict summaries, but they should not be interpreted as direct source-specific borrowing amounts.

\noindent\textit{Unit information prior.}
For the UIP, posterior summaries of \(M w_{{\rm UIP},k}\) can be interpreted as source-specific contributions to the information incorporated through the constructed prior.
However, they should be distinguished from direct source-specific borrowing amounts, because the UIP combines historical summaries into a single prior distribution before updating with the current control data.

\noindent\textit{Multisource exchangeability model.}
For the MEM, source-specific borrowing can be summarized by the posterior probability that the $k$-th historical control is exchangeable with the current control.

\noindent\textit{Dirichlet process mixture models.}
For the DPM and DDPM models, source-specific borrowing and compatibility are summarized by the SBI.
For the $k$-th historical control, ${\rm SBI}_k$ is defined as the posterior probability that the $k$-th historical control and the current control are assigned to the same mixture component.
Because shared clustering implies that the two data sources are represented by the same parameter value, a larger ${\rm SBI}_k$ indicates stronger posterior support for both similarity and borrowing.
Unlike EHSS, the SBI is not measured on the sample-size scale; rather, it is a posterior probability that summarizes source-specific support for borrowing through the clustering structure.

Overall borrowing measures and source-specific summaries provide complementary information.
The EHSS provides a scalar summary of the total amount of information borrowed from historical controls on the sample-size scale.
In contrast, source-specific summaries identify which historical controls are likely to contribute to borrowing or be discounted because of incompatibility with the current control.
Because source-specific summaries are method-specific, they are not always directly comparable on the same numerical scale.
For example, the power parameter, exchangeability probability, SBI, and potential bias parameter represent different aspects of borrowing and compatibility.

In the case studies presented below, we use overall borrowing summaries and method-specific source-level summaries to compare how dynamic borrowing methods respond to prior--data conflict.

\section{Case study 1: Binary endpoint example\label{sec:binary}}

\subsection{Data and methods}

We illustrate the posterior quantification of borrowing using a clinical trial example with a binary endpoint.
This example is based on a phase II proof-of-concept trial of secukinumab for the treatment of ankylosing spondylitis \citep{baetenAntiinterleukin17AMonoclonalAntibody2013}.
The endpoint was achievement of a 20\% response according to the SpondyloArthritis International Society criteria for improvement.
The data consist of eight historical control studies and one current randomized trial.
Table~\ref{tab:case_binary_data} presents the sample sizes, numbers of responses, and response rates for the historical controls and the current trial.

The treatment effect was defined as the difference in response probabilities between the treatment and current control arms,
$p_{\rm CT}-p_{\rm CC}$.
We compared the MAP, DPM-MAP, DMPP, UIP, PBM with the horseshoe prior (HS), MEM, DPM, and DDPM methods.
As a reference analysis, we also included a current-only analysis that used only the current trial data without borrowing from historical controls.
For each method, we summarized the posterior mean and 95\% credible interval of the treatment effect.
For all methods, the treatment-arm response probability was assigned a common uniform prior,
$p_{\rm CT} \sim {\rm Beta}(1,1)$.
For the MAP analysis, we used the \texttt{RBesT} package \citep{weberApplyingMetaAnalyticPredictivePriors2021} with a half-normal prior with scale 1 for the between-study heterogeneity parameter and a weakly informative normal prior for the overall mean parameter.
The resulting MAP prior was approximated by a three-component beta mixture and updated with the current control data.
For the DPM-MAP analysis, we used a finite stick-breaking approximation with 10 mixture components, beta priors for the stick-breaking variables, half-normal priors with scale 1 for the component-specific standard deviations, and a weakly informative normal prior for the overall mean parameter.
For the DMPP analysis, the power parameters were modeled hierarchically as
$\gamma_{{\rm H}_k} \sim {\rm Beta}(\mu_{\rm PP}\kappa_{\rm PP}, (1-\mu_{\rm PP})\kappa_{\rm PP})$,
with $\mu_{\rm PP} \sim {\rm Beta}(1,1)$ and
$\kappa_{\rm PP} \sim {\rm Lognormal}(\log2,1)$.
This prior centers the beta precision around $\kappa_{\rm PP}=2$ while allowing substantial prior variability in the dispersion of the source-specific power parameters.
For the UIP analysis, the weight vector was assigned a Dirichlet prior with parameters $\nu_{{\rm H}_k}=\min(1,n_{{\rm H}_k}/n_{\rm CC})$, and the total amount of information $M$ was assigned a uniform prior on
$\left(0,\sum_{k=1}^K n_{{\rm H}_k}\right)$.
For the HS analysis, the source-specific potential bias parameters were assigned horseshoe priors, with half-Cauchy priors for the global and local shrinkage parameters.
For the MEM analysis, all possible exchangeability patterns between the current and historical controls were enumerated.
A beta-binomial model with ${\rm Beta}(1,1)$ priors was used for the response probabilities, and the prior probabilities of exchangeability patterns were induced by a ${\rm Beta}(1,1)$ prior on the inclusion probability.
For the DPM and DDPM analyses, the concentration parameter of the Dirichlet process was assigned a gamma prior with shape 1 and scale 5.

Overall borrowing was quantified using the EHSS, defined as the posterior ESS for the current control parameter minus the current control sample size.
The posterior ESS was calculated using the expected local-information-ratio (ELIR) method implemented in the \texttt{RBesT} package \citep{weberApplyingMetaAnalyticPredictivePriors2021}.
For methods with source-specific summaries, we also examined posterior source-specific borrowing or compatibility summaries for each historical study.

\begin{table}[tbp]
\centering
\caption{Observed sample sizes, numbers of responses, and response rates in the ankylosing spondylitis example.}
\label{tab:case_binary_data}
\begin{tabular}{lrrrrrrrrrrr}
\toprule
& \multicolumn{9}{c}{Historical controls} & \multicolumn{2}{c}{Current trial} \\
\cmidrule(lr){2-10}\cmidrule(lr){11-12}
Variable
& $\text{H}_1$ & $\text{H}_2$ & $\text{H}_3$ & $\text{H}_4$ & $\text{H}_5$ & $\text{H}_6$ & $\text{H}_7$ & $\text{H}_8$ & Sum
& CC & CT \\
\midrule
Response
& 23 & 12 & 19 & 9 & 39 & 6 & 9 & 10 & 127
& 1 & 14 \\
$N$
& 107 & 44 & 51 & 39 & 139 & 20 & 78 & 35 & 513
& 6 & 23 \\
Response rate (\%)
& 21.5 & 27.3 & 37.3 & 23.1 & 28.1 & 30.0 & 11.5 & 28.6 & 24.8
& 16.7 & 60.9 \\
\bottomrule
\end{tabular}
\end{table}

\subsection{Results}

\begin{figure}[htbp]
\centering
\includegraphics[width=0.85\textwidth]{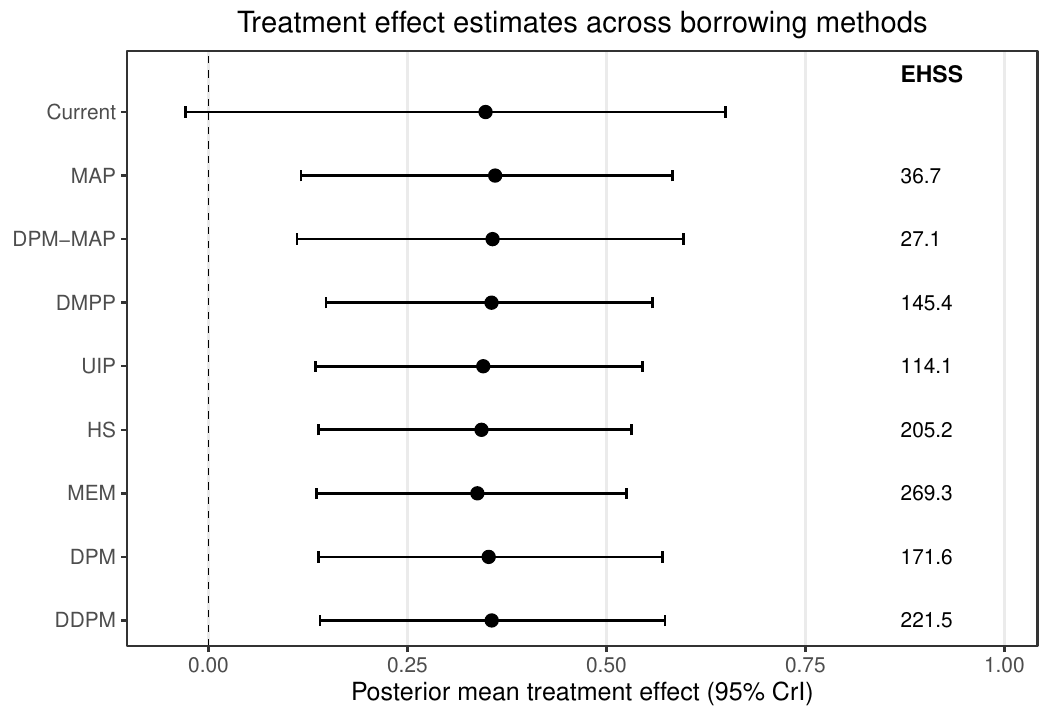}
\caption{Posterior treatment effect estimates and effective historical sample size (EHSS) for the binary endpoint example. Points and horizontal bars represent posterior means and 95\% credible intervals for the treatment effect, respectively. EHSS values summarize the overall amount of information borrowed from historical controls.}
\label{fig:forest_binary}
\end{figure}

Figure~\ref{fig:forest_binary} shows the posterior treatment effect estimates for the current-only and borrowing analyses, together with EHSS values for the borrowing analyses.
The EHSS was not reported for the current-only analysis because no historical information was borrowed.
All methods yielded positive posterior mean treatment effects, and the 95\% credible intervals were broadly similar across methods.
Based on these results, the posterior treatment effect estimates alone suggested qualitatively similar conclusions.
In contrast, the EHSS varied substantially across methods.
The EHSS ranged from 27.1 for DPM-MAP to 269.3 for MEM.
This indicates that different dynamic borrowing methods can yield similar posterior treatment effect estimates while borrowing substantially different amounts of information from historical controls.
Therefore, EHSS provides information about the overall contribution of historical data that is not apparent from the treatment effect estimate alone.

To examine how borrowing was distributed across individual historical sources, we considered source-specific posterior summaries and displayed them in a heatmap.
The heatmap was restricted to DMPP, MEM, DPM, and DDPM, because these methods provide source-specific posterior summaries on a 0--1 scale with a direct borrowing or compatibility interpretation.
MAP and DPM-MAP do not define source-specific borrowing parameters.
UIP was not included because $M w_{{\rm UIP},k}$ summarizes the posterior source-specific contribution to the information incorporated through the constructed UIP, rather than a direct 0--1 posterior borrowing or compatibility summary.
HS provides posterior summaries of source-specific conflict rather than direct borrowing probabilities.

\begin{figure}[htbp]
\centering
\includegraphics[width=0.85\textwidth]{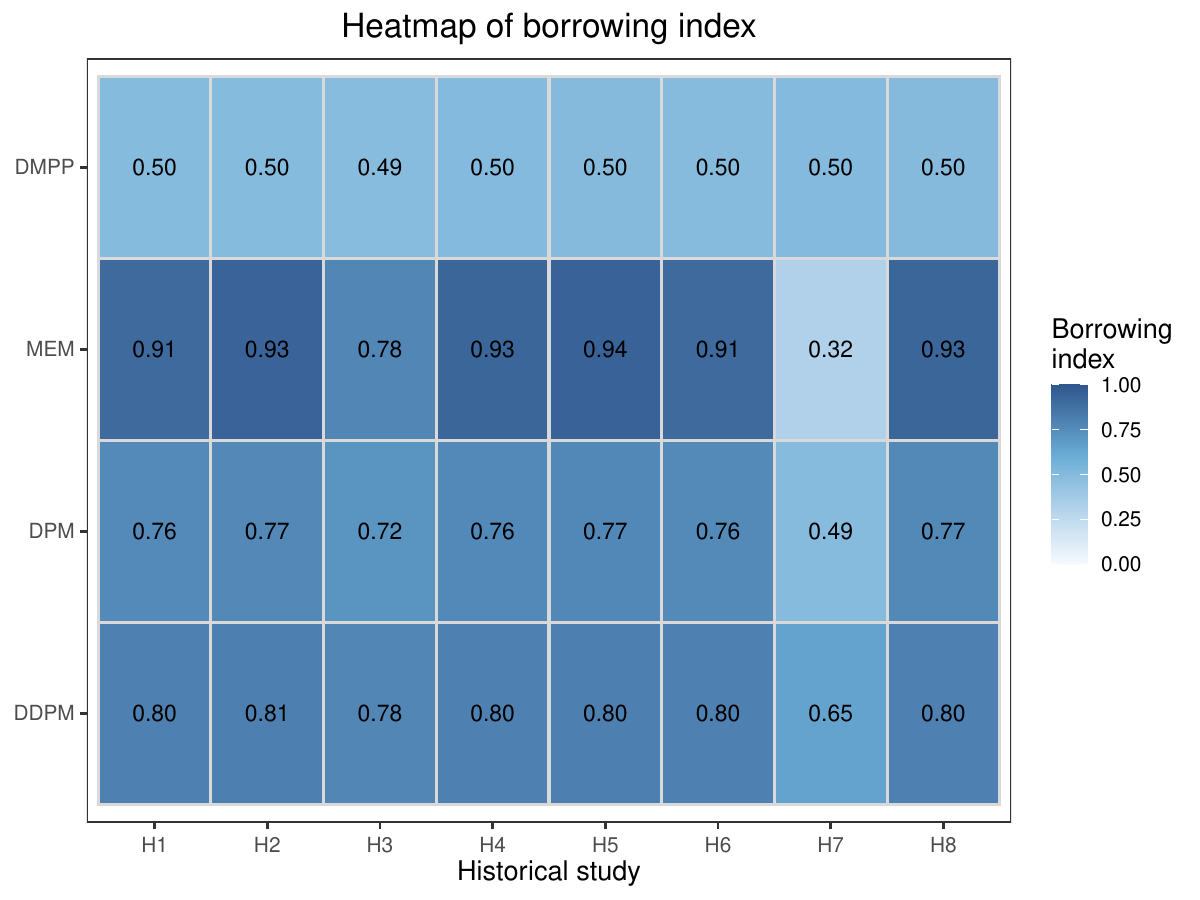}
\caption{Source-specific borrowing or compatibility summaries (borrowing index) for the binary endpoint example. The color scale labeled ``source-specific summary'' represents method-specific posterior quantities on a 0--1 scale. For DMPP, values are posterior means of the source-specific power parameters. For MEM, values are posterior exchangeability probabilities. For DPM and DDPM, values are borrowing indices based on posterior shared clustering.}
\label{fig:heat_binary}
\end{figure}

Figure~\ref{fig:heat_binary} shows the source-specific borrowing or compatibility summaries for DMPP, MEM, DPM, and DDPM, collectively referred to as the borrowing index.
Although these quantities are displayed on a common 0--1 scale, their interpretations differ by method: DMPP values are posterior means of the source-specific power parameters, MEM values are posterior exchangeability probabilities, and DPM/DDPM values are SBIs based on posterior shared clustering.
For DMPP, the posterior mean power parameters were close to 0.5 for all historical studies, indicating relatively uniform likelihood discounting across sources.
This pattern suggests that, in this binary example with a very small current control arm, the source-specific power parameters were only weakly informed by the data and remained close to the prior center.
In contrast, MEM, DPM, and DDPM showed more heterogeneous source-specific patterns.
For MEM, the posterior exchangeability probabilities were high for most historical studies but were substantially lower for $\text{H}_7$.
Similarly, the SBI values for DPM and DDPM were lower for $\text{H}_7$ than for the other historical studies.
These patterns suggest that $\text{H}_7$ was assessed as less compatible with the current control than the other historical controls under methods that allow source-specific exchangeability or clustering.

\section{Case study 2: Continuous endpoint example\label{sec:cont}}

\subsection{Data and methods}

We next illustrate posterior quantification of borrowing using a clinical trial example with a continuous endpoint.
This example is based on the Alzheimer's Disease Cooperative Study (ADCS) dataset analyzed by \citet{banbetaPowerPriorMultiple2022}.
The current trial was ADC-037, which evaluated resveratrol in individuals with mild to moderate Alzheimer's disease.
The outcome was the change in Alzheimer's Disease Assessment Scale--cognitive subscale (ADAS-cog) score from baseline to week 52.
Because a larger increase in ADAS-cog indicates greater cognitive decline, a negative treatment effect favors the treatment arm.
The data consist of five historical control studies and one current randomized trial.
Table~\ref{tab:case_cont_data} presents the sample sizes, mean changes, and standard deviations of the change in the ADAS-cog score for the historical controls and the current trial.
For this example, we used arm-level summary statistics for the change score and did not adjust for baseline covariates.

\begin{table}[tbp]
\centering
\caption{Observed sample sizes, mean changes, and standard deviations of the change in ADAS-cog score within the ADCS dataset.}
\label{tab:case_cont_data}
\begin{tabular}{lrrrrrrrr}
\toprule
& \multicolumn{5}{c}{Historical controls} & \multicolumn{2}{c}{Current trial} \\
\cmidrule(lr){2-6}\cmidrule(lr){7-8}
Variable
& $\text{H}_1$ & $\text{H}_2$ & $\text{H}_3$ & $\text{H}_4$ & $\text{H}_5$
& CC & CT \\
\midrule
Trial
& ADC-11 & ADC-15 & ADC-16 & ADC-22 & ADC-27
& ADC-37 & ADC-37 \\
$N$
& 111 & 202 & 169 & 63 & 164
& 55 & 64 \\
Mean change
& 8.7 & 8.0 & 4.4 & 6.1 & 5.1
& 4.8 & 2.9 \\
SD
& 7.2 & 5.8 & 6.4 & 7.1 & 6.8
& 6.3 & 9.6 \\
\bottomrule
\end{tabular}
\end{table}

The treatment effect was defined as the difference in mean change scores between the treatment and current control arms,
$\theta_{\rm CT}-\theta_{\rm CC}$.
We compared the MAP, DPM-MAP, DMPP, UIP, PBM with the horseshoe prior (HS), MEM, DPM, and DDPM methods.
As in the binary endpoint example, a current-only analysis was included as a reference.
For each method, we summarized the posterior mean and 95\% credible interval of the treatment effect.
For all methods, the treatment-arm mean was assigned a common weakly informative normal prior,
$\theta_{\rm CT} \sim \text{N}(0,100^2)$.
Given the observed treatment-arm mean $\bar{y}_{\rm CT}$ and standard error ${\rm SE}_{\rm CT}$, the sampling model was
$\bar{y}_{\rm CT} \mid \theta_{\rm CT} \sim \text{N}(\theta_{\rm CT},{\rm SE}_{\rm CT}^2)$.
The borrowing-model prior and hyperparameter settings were the same as those used in the binary endpoint example, except for settings specific to the continuous endpoint.
For the MAP analysis, we used the \texttt{RBesT} package \citep{weberApplyingMetaAnalyticPredictivePriors2021} with a half-normal prior with scale $6.77/2$ for the between-study heterogeneity parameter and a weakly informative normal prior for the overall mean parameter.
A value of 6.77 was used as a representative standard deviation for the change in ADAS-cog score, following the standard deviation reported from the historical ADCS trials.
The resulting MAP prior was approximated by a three-component normal mixture and updated with the current control data.
For the DPM-MAP analysis, we used the same finite stick-breaking approximation as in the binary endpoint example, with normal likelihoods for the arm-level mean changes.
For the UIP analysis, the weight vector was assigned a Dirichlet prior with parameters equal to 1, and the total amount of information $M$ was assigned a uniform prior on
$\left(0,\sum_{k=1}^K n_{{\rm H}_k}\right)$.
The DMPP, HS, MEM, DPM, and DDPM analyses used the same corresponding hyperparameter settings as in the binary endpoint example.

Overall borrowing was quantified using the EHSS, defined as the posterior ESS for the current control parameter minus the current control sample size.
The posterior ESS was calculated using the expected local-information-ratio (ELIR) method implemented in the \texttt{RBesT} package \citep{weberApplyingMetaAnalyticPredictivePriors2021}.
For the continuous endpoint, the reference standard deviation for the ELIR calculation was set to the observed standard deviation of the current control arm, 6.3, so that EHSS was expressed on the information scale of the current control data.
For methods with source-specific summaries, we also examined posterior source-specific borrowing or compatibility summaries for each historical study.

\subsection{Results}

\begin{figure}[htbp]
\centering
\includegraphics[width=0.85\textwidth]{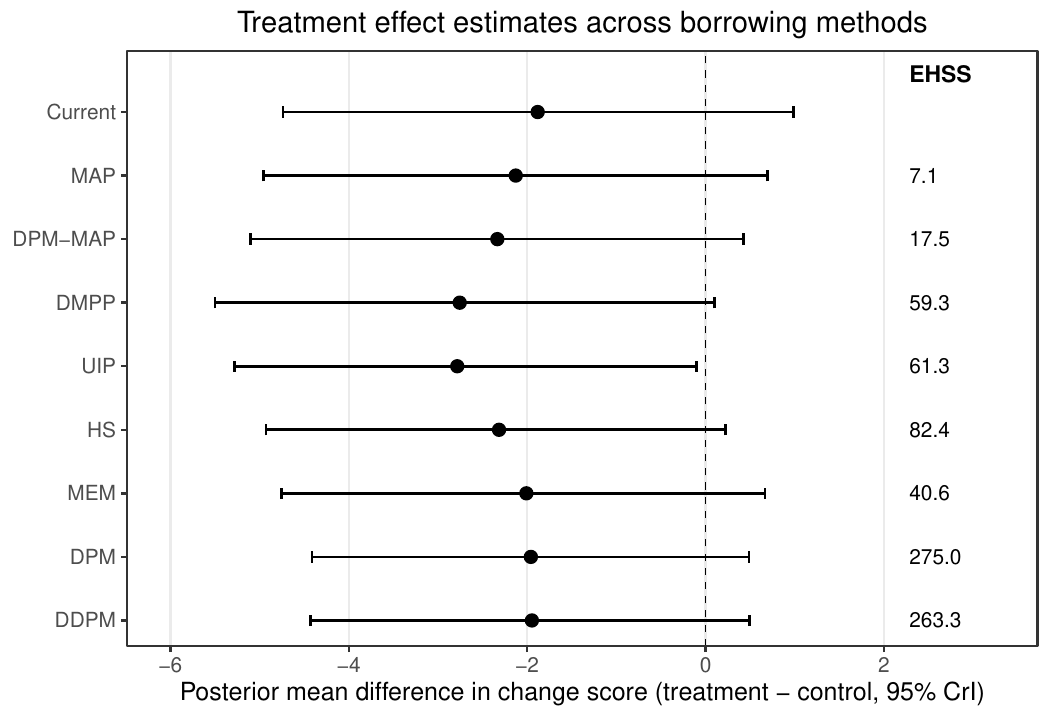}
\caption{Posterior treatment effect estimates and effective historical sample size (EHSS) for the continuous endpoint example. Points and horizontal bars represent posterior means and 95\% credible intervals for the treatment effect, respectively. Negative values favor treatment because larger increases in ADAS-cog score indicate greater cognitive decline. EHSS values summarize the overall amount of information borrowed from historical controls.}
\label{fig:forest_cont}
\end{figure}

Figure~\ref{fig:forest_cont} shows the posterior treatment effect estimates for the current-only and borrowing analyses, together with EHSS values for the borrowing analyses.
All methods yielded negative posterior mean treatment effects, indicating a smaller increase in ADAS-cog score in the treatment arm than in the current control arm.
However, the 95\% credible intervals were wide and included zero for all methods.
Thus, the posterior treatment effect estimates suggested similar qualitative conclusions across methods.
In contrast, the EHSS varied substantially across methods.
The EHSS ranged from 9.4 for MAP to 259.3 for DPM.
The MAP, DPM-MAP, DMPP, UIP, HS, and MEM methods yielded small-to-moderate EHSS values, whereas DPM and DDPM yielded substantially larger EHSS values.
This indicates that, even when posterior treatment effect estimates are broadly similar, the amount of information borrowed from historical controls can differ markedly across methods.

\begin{figure}[htbp]
\centering
\includegraphics[width=0.85\textwidth]{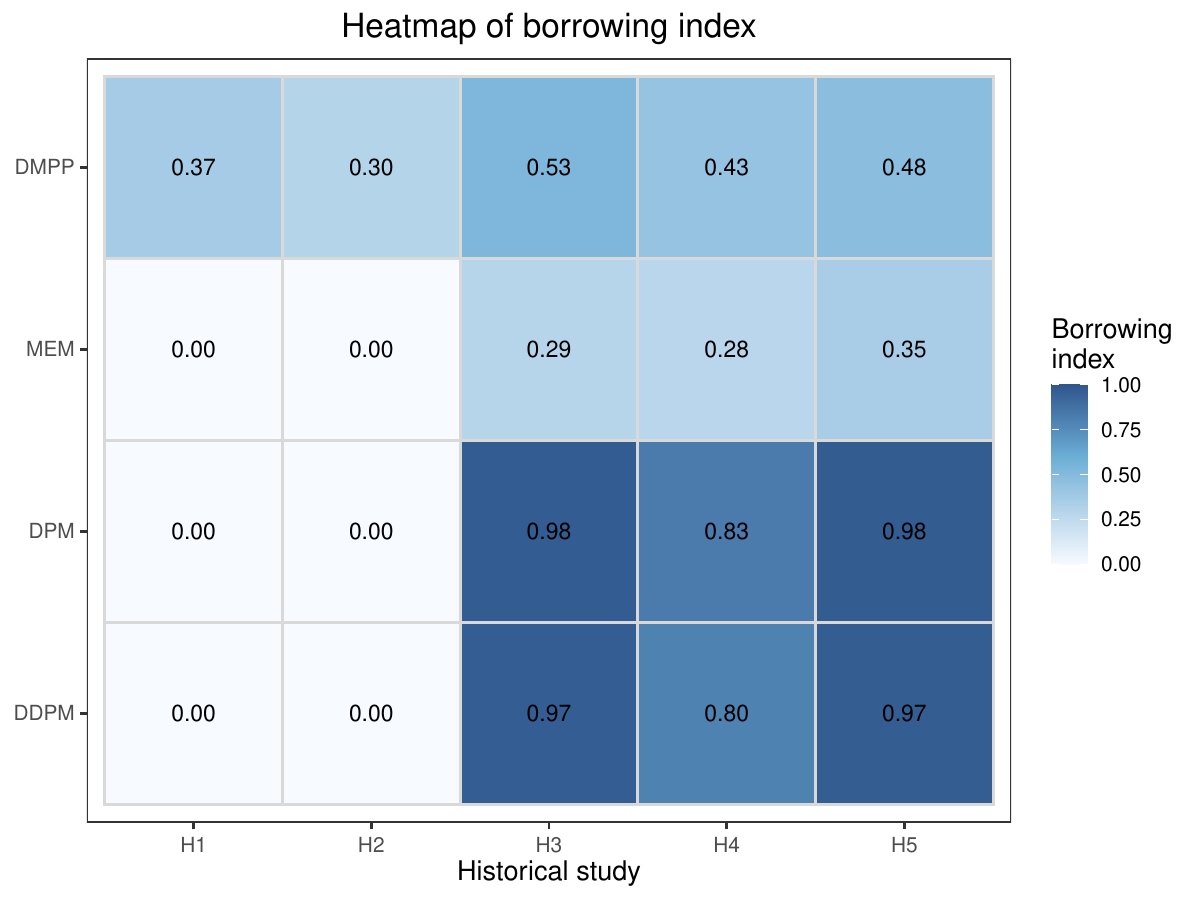}
\caption{Source-specific borrowing or compatibility summaries (borrowing index) for the continuous endpoint example. The color scale labeled ``source-specific summary'' represents method-specific posterior quantities on a 0--1 scale. For DMPP, values are posterior means of the source-specific power parameters. For MEM, values are posterior exchangeability probabilities. For DPM and DDPM, values are borrowing indices based on posterior shared clustering.}
\label{fig:heat_cont}
\end{figure}

Using the same criterion as in the binary endpoint example, we displayed source-specific posterior summaries for DMPP, MEM, DPM, and DDPM in a heatmap.
Figure~\ref{fig:heat_cont} shows the source-specific borrowing or compatibility summaries for DMPP, MEM, DPM, and DDPM, collectively referred to as the borrowing index.
For DMPP, the posterior mean power parameters varied across historical studies, with lower values for $\text{H}_1$ and $\text{H}_2$ and higher values for $\text{H}_3$ and $\text{H}_5$.
For MEM, the posterior exchangeability probabilities were close to zero for $\text{H}_1$ and $\text{H}_2$, but were more moderate for $\text{H}_3$, $\text{H}_4$, and $\text{H}_5$.
In contrast, the SBI values for DPM and DDPM were close to zero for $\text{H}_1$ and $\text{H}_2$ and very high for $\text{H}_3$, $\text{H}_4$, and $\text{H}_5$.
This pattern is consistent with the observed mean changes: $\text{H}_1$ and $\text{H}_2$ showed larger mean increases in ADAS-cog than the current control, whereas $\text{H}_3$ and $\text{H}_5$ were closer to the current control.
Overall, methods based on exchangeability or clustering assigned less support for borrowing from $\text{H}_1$ and $\text{H}_2$, although the degree of support for borrowing from $\text{H}_3$, $\text{H}_4$, and $\text{H}_5$ differed across methods.

Overall, this case study suggests that source-specific summaries can reveal which historical controls drive the overall borrowing.
In this example, the large EHSS values for DPM and DDPM were accompanied by high SBIs for $\text{H}_3$, $\text{H}_4$, and $\text{H}_5$ but almost no borrowing from $\text{H}_1$ and $\text{H}_2$.
Thus, large overall borrowing did not imply uniform borrowing from all historical controls; rather, it reflected selective borrowing from historical controls assessed as compatible with the current control.

\section{Practical recommendations for reporting and interpretation\label{sec:recomm}}  

The case studies highlight that posterior treatment effect estimates alone are insufficient for understanding how Bayesian dynamic borrowing methods use historical information.
Different methods can yield broadly similar posterior treatment effect estimates while borrowing different amounts of information from historical controls.
Therefore, borrowing should be reported explicitly as part of the interpretation of a dynamic borrowing analysis.

First, overall borrowing should be summarized using an interpretable scalar measure such as EHSS \citep{hobbsAdaptiveAdjustmentRandomization2013,wiesenfarthQuantificationPriorImpact2020,bennettNovelEquivalenceProbability2021}.
Because EHSS is expressed on the sample-size scale, it provides a direct summary of how much additional information is contributed by the historical controls beyond the current control data.
However, EHSS should not be interpreted in isolation.
A large EHSS indicates substantial overall borrowing, but it cannot identify which historical sources drive the borrowing or whether within-conflict is present for particular sources.

Second, when multiple historical control sources are used, source-specific summaries should be reported whenever they are available.
These summaries can reveal whether borrowing is supported broadly across historical controls or concentrated in a subset of compatible sources.
For example, in the continuous endpoint example, the DPM and DDPM methods yielded large EHSS values; however, the corresponding SBIs indicated that borrowing was concentrated in historical controls that were more compatible with the current control.
Thus, source-specific summaries provide information that cannot be obtained from EHSS alone.

Third, source-specific summaries should be interpreted according to the borrowing mechanism of each method.
For power prior methods, the posterior mean or credible interval of the source-specific power parameter directly summarizes the contribution of the corresponding historical likelihood.
For MEM, the posterior exchangeability probability summarizes posterior support that a historical control is exchangeable with the current control.
For DPM and DDPM, the SBI summarizes posterior support for shared clustering between the historical and current control parameters.
For PBMs, posterior summaries of the potential bias and shrinkage parameters characterize source-specific compatibility or conflict; however, they should not be interpreted as direct borrowing amounts.
For UIP, source-specific quantities such as \(M w_{{\rm UIP},k}\) summarize the posterior allocation of information through the constructed UIP; however, they should be distinguished from direct likelihood-based borrowing parameters.

Finally, borrowing summaries should be interpreted together with clinical and design
information about the historical controls
\citep{vieleUseHistoricalControl2014,lesaffreReviewDynamicBorrowing2024}.
A low source-specific borrowing summary may indicate within-conflict, but it may also reflect limited information in the current data, differences in model assumptions, or uncertainty in the compatibility assessment.
Conversely, a high borrowing summary should not be interpreted as evidence that the
historical source is clinically exchangeable unless supported by trial design, eligibility
criteria, endpoint definitions, and other external considerations.
Therefore, posterior borrowing summaries should complement, rather than replace,
prespecified assessments of the relevance of historical data.

\section{Discussion\label{sec:dis}}

This review focused on how the degree of information borrowing can be quantified and interpreted after fitting Bayesian dynamic borrowing models with multiple historical controls.
The two case studies showed that methods with similar posterior treatment effect estimates can differ in both overall borrowing and the distribution of borrowing across historical sources, thereby providing different perspectives on between-conflict and within-conflict.

A key challenge is that source-specific borrowing is not equally well defined across methods.
For power prior methods, MEM, DPM, and DDPM, source-specific posterior summaries have relatively direct interpretations through power parameters, exchangeability probabilities, or shared-clustering probabilities.
In contrast, MAP-type approaches, including robust MAP and DPM-MAP \citep{neuenschwanderSummarizingHistoricalInformation2010,schmidliRobustMetaanalyticpredictivePriors2014,hupfBayesianSemiparametricMetaanalyticpredictive2021}, primarily provide global borrowing summaries.
Because the MAP prior is constructed in a two-stage manner from the collection of historical controls, the posterior analysis does not naturally retain source-specific borrowing parameters.
This limitation is not unique to two-stage MAP analyses; even in one-stage meta-analytic combined analyses, borrowing is typically induced through a common hierarchical model and may not be directly decomposable into source-specific borrowing amounts.
Related work on Bayesian hierarchical models, such as the individual borrowing strength and overall borrowing index proposed by \citet{xuBorrowingStrengthBorrowing2020}, may provide a basis for future development.
However, for two-stage MAP analyses, any source-specific contribution may be more naturally interpreted as the contribution of each historical source to the constructed prior, rather than as a posterior source-specific borrowing amount after observing the current control data.

The interpretation of source-specific borrowing is also challenging for PBMs.
In PBMs, the potential bias parameter $\beta_{{\rm H}_k}$ directly represents conflict or compatibility between the current control and the $k$-th historical control.
However, translating this conflict summary into a borrowing amount is not straightforward because borrowing is induced indirectly through the shrinkage of the potential bias parameters rather than through explicit likelihood discounting, exchangeability indicators, or shared-clustering probabilities.
Thus, posterior summaries of $\beta_{{\rm H}_k}$ and the corresponding shrinkage parameters should be interpreted primarily as source-specific conflict or compatibility summaries, not as direct source-specific borrowing amounts.
For PBMs with shrinkage priors, concepts such as the effective number of nonzero coefficients proposed by \citet{piironenSparsityInformationRegularization2017} may be useful for summarizing the effective number of historical controls that are treated as conflicting with the current control.

In conclusion, posterior quantification of borrowing provides an important layer of transparency in Bayesian dynamic borrowing analyses with multiple historical controls.
Overall borrowing summaries, source-specific posterior summaries, and clinical assessments of the relevance of historical data should be reported together.
This combined reporting can help investigators understand not only the final treatment effect estimate, but also how and from which historical sources that estimate was informed.

%---------------------------------------%
%          Acknowledgement              %
%---------------------------------------%
\section*{Acknowledgement}
This work was supported by the JSPS KAKENHI Grant Numbers 24K20739, and 24K02662.

%---------------------------------------------%
%                Reference                    %
%---------------------------------------------%

\end{document}